\documentclass[prl,twocolumn,superscriptaddress,showpacs]{revtex4-1}
\usepackage{graphicx,amsmath,amssymb,bm}

\newcommand{\be}{\begin{equation}}	
\newcommand{\ee}{\end{equation}}
\newcommand{\vlowk}{V_{{\rm low}\,k}}
\newcommand{\lm}{\Lambda}
\newcommand{\fmi}{\, \text{fm}^{-1}}
\newcommand{\mev}{\, \text{MeV}}

\begin{document}

\title{Three-body forces and shell structure in calcium isotopes}

\author{Jason D.\ Holt}
\affiliation{Department of Physics and Astronomy, University of Tennessee, 
Knoxville, TN 37996, USA}
\affiliation{Physics Division, Oak Ridge National Laboratory, P.O. Box 2008,
Oak Ridge, TN 37831, USA}
\author{Takaharu Otsuka}
\affiliation{Department of Physics and Center for Nuclear Study,
University of Tokyo, Hongo, Tokyo 113-0033, Japan}
\affiliation{National Superconducting Cyclotron Laboratory,
Michigan State University, East Lansing, MI, 48824, USA}
\author{Achim Schwenk}
\affiliation{ExtreMe Matter Institute EMMI, GSI Helmholtzzentrum f\"ur
Schwerionenforschung GmbH, 64291 Darmstadt, Germany}
\affiliation{Institut f\"ur Kernphysik, Technische Universit\"at
Darmstadt, 64289 Darmstadt, Germany}
\author{Toshio Suzuki}
\affiliation{Department of Physics, Nihon University, Sakurajosui 3, 
Tokyo 156-8550, Japan}

\begin{abstract}
Understanding and predicting the formation of shell structure from
nuclear forces is a central challenge for nuclear physics. While the magic
numbers $N=2,8,20$ are generally well understood, $N=28$ is the
first standard magic number that is not reproduced in microscopic
theories with two-nucleon forces. In this Letter, we show that
three-nucleon forces give rise to repulsive interactions between
two valence neutrons that are key to explain $^{48}$Ca as a magic nucleus,
with a high $2^+$ excitation energy and a concentrated magnetic
dipole transition strength. The repulsive
three-nucleon mechanism improves the agreement with experimental
binding energies.
\end{abstract}


\maketitle

In nuclei certain configurations of protons and neutrons (nucleons)
are observed to be particularly well-bound. These closed-shell or
``magic" nuclei form the basis of the nuclear shell model (exact
diagonalizations in spaces based on the observed shell
structure~\cite{RMP}), which provides a key computational method in
nuclear physics. Exploring the formation of shell structure and how
these magic configurations evolve with nucleon number towards the drip
lines is a frontier in the physics of nuclei, and a microscopic
understanding from nuclear forces presents a major challenge for
theory.

\begin{figure*}
\begin{center}
\includegraphics[scale=0.55,clip=]{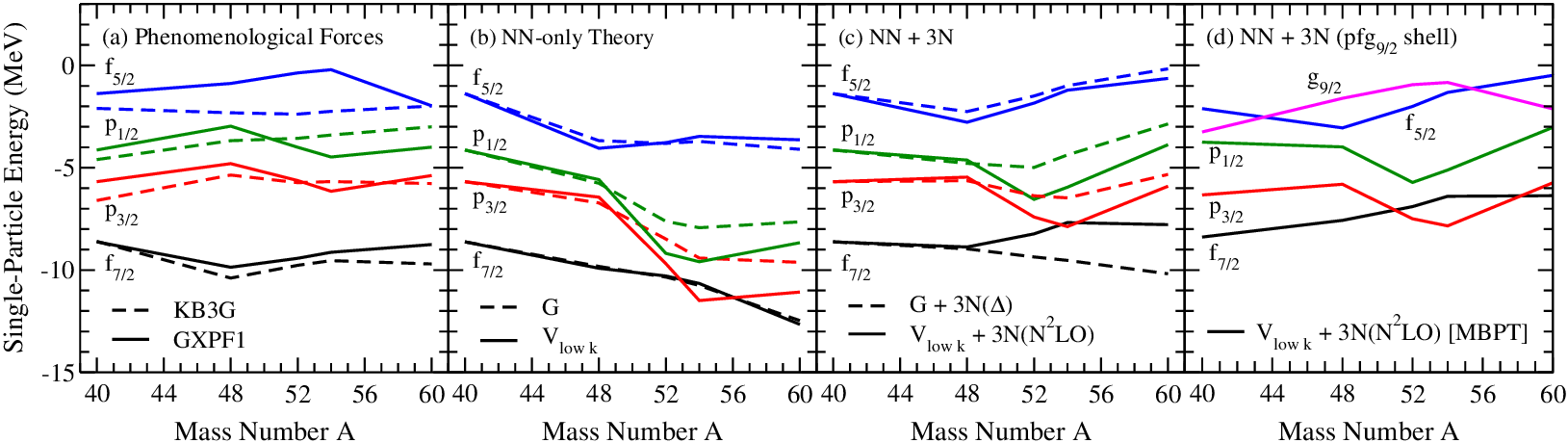}
\end{center}
\vspace*{-5mm}
\caption{Neutron SPEs relative to the $^{40}$Ca
energy as a function of mass number $A$. (a)~SPEs obtained from
phenomenological models
KB3G~\cite{KB3G} and GXPF1~\cite{GXPF1}. 
(b)~NN-only theory: SPEs calculated from a 
$G$-matrix and low-momentum interactions $\vlowk$. (c)~SPEs including
contributions from 3N forces due to $\Delta$ excitations,
3N($\Delta$), and chiral EFT 3N interactions at N$^2$LO,
3N(N$^2$LO)~\cite{3Nfit}. The results in panels~(b) and~(c) start
from the empirical GXPF1 SPEs in $^{41}$Ca. (d)~SPEs in the
$pfg_{9/2}$ shell calculated from $\vlowk$ and 3N(N$^2$LO) forces,
where the starting SPEs in $^{41}$Ca are calculated consistently in
many-body perturbation theory [MBPT] 
and include one-body 3N contributions.\label{spe}}
\vspace*{-4mm}
\end{figure*}

The theoretical shortcomings in predicting shell structure are
particularly evident in the calcium isotopes. While microscopic
calculations with well-established two-nucleon (NN) forces reproduce
the standard magic numbers $N=2, 8, 20$, one of the most striking
failures is that they do not predict $^{48}$Ca as a doubly-magic
nucleus when neutrons are added to $^{40}$Ca~\cite{PovesZuker,RMP},
making $N=28$ the first standard magic number not reproduced in
microscopic NN theories. As a result, phenomenological forces have
been adjusted to yield a doubly-magic $^{48}$Ca~\cite{GXPF1,KB3G}, and
it has been argued these phenomenological adjustments may be largely
due to neglected three-nucleon (3N) forces~\cite{Zuker}. Recently, we
have shown that 3N forces play a decisive role for the oxygen anomaly
and can explain why $^{24}$O is the heaviest oxygen
isotope~\cite{Oxygen}. In this Letter, we present the first study of
the impact of 3N forces on medium-mass nuclei. Our results demonstrate
that one- and two-body contributions from 3N forces to valence
neutrons, as well as extended valence spaces,
are essential to understand shell structure in the calcium
isotopes and $N=28$ as a magic number based on nuclear forces.

Three-nucleon forces were introduced in the pioneering work of Fujita
and Miyazawa (FM)~\cite{FM} and arise because nucleons are finite-mass
composite particles that can also be excited by interacting with other
particles. The long-range part of 3N forces is dominated by the FM 3N
mechanism, where one nucleon virtually excites a second nucleon to the
$\Delta(1232 \mev)$ resonance, which is de-excited by interacting with
a third nucleon. Additional long-range and shorter-range 3N
interactions are included naturally in chiral effective field theory
(EFT)~\cite{chiral}, which provides a systematic expansion for nuclear
forces. The importance of chiral 3N forces has been well
established in light nuclei with $A=N+Z \lesssim 12$~\cite{light}.

We derive the interactions among valence neutrons following two
approaches. First, we use low-momentum interactions $\vlowk$ with
smooth cutoffs~\cite{Vlowk} obtained by evolving a chiral N$^3$LO NN
potential~\cite{N3LO} to lower resolution with $\lm = 2.0 \fmi$.  The
two-body interactions in the $pf$ and $pfg_{9/2}$ shell are calculated
to third order in many-body perturbation theory (MBPT) following
Refs.~\cite{LNP,Gmatrix} in a space of $13$ major shells. We use a
harmonic oscillator basis with $\hbar \omega = 11.48 \mev$,
appropriate for the calcium isotopes. Our results are converged for
intermediate-state excitations to $18 \hbar\omega$.  Second, we take a
standard $G$-matrix that has been used as a starting point in many
nuclear-structure calculations~\cite{Gmatrix}. This $G$ matrix is
based on the Bonn~C NN potential and includes many-body contributions
to third order, but with intermediate-state excitations to $2 \hbar
\omega$ (and calculated for $\hbar \omega = 10 \mev$).
Although the $G$-matrix calculation should be improved by
extending the intermediate states, this two-body interaction has
been used as standard starting point for shell-model studies, and
gives us a baseline to investigate changes due to three-body forces.
A detailed study of the cutoff dependence, which provides a measure of
the theoretical uncertainty, is left to future work.

To study the validity of the MBPT approach, we have carried out
coupled-cluster (CC) calculations for the ground-state energies of
$^{40,48,52,54}$Ca at the $\Lambda$-CCSD(T) level~\cite{CClong}, based
on the same $\vlowk$ interaction and basis space. We have verified
that the CC energies are converged in these spaces. Using
particle-attached (to $^{40}$Ca) CC energies as single-particle
energies (SPEs), the MBPT results agree within a few percent with CC
theory (relative to $^{40}$Ca): -$159.3 \mev$ (MBPT) vs.~-$155.0 \mev$
(CC) for $^{48}$Ca; -$230.7 \mev$ vs.~-$235.9 \mev$ for $^{52}$Ca; and
-$259.1 \mev$ vs.~-$268.6 \mev$ for $^{54}$Ca. This shows that MBPT
can be comparable to CC theory for $\vlowk$ interactions, but also
highlights the important role of SPEs.

To understand shell structure in the calcium isotopes, we show in
Fig.~\ref{spe} the change of the SPEs of the neutron orbitals, with
standard quantum numbers $l_j$, in the $pf$ and $pfg_{9/2}$ shell as
neutrons are added to $^{40}$Ca. Fig.~\ref{spe}~(a) gives the
evolution obtained from phenomenological models: GXPF1~\cite{GXPF1}, a
quasi-global fit of two-body matrix elements (starting from a
G-matrix) and of initial SPEs in $^{41}$Ca to experimental data; and
KB3G~\cite{KB3G}, which modifies the monopole part of a standard $G$
matrix. Despite different initial SPEs,
as neutrons fill the $f_{7/2}$ shell, the repulsive interaction
between the $f_{7/2}$ and $p_{3/2}$ neutrons causes a significant gap
to develop at $N=28$, indicative of a shell closure.

In contrast, both NN-only theories in Fig.~\ref{spe}~(b) exhibit
minimal repulsion between these orbitals and the gap remains largely
unchanged from $^{40}$Ca. This is even more evident when the $^{41}$Ca
SPEs are calculated microscopically at the same third-order level: the
NN-only SPEs (based on $\vlowk$) are too bound and both $f_{7/2}$ and
$p_{3/2}$ orbitals are at $-10.8 \mev$. Starting from the GXPF1 SPEs
in $^{41}$Ca, the NN-only results in Fig.~\ref{spe}~(b) depend only
weakly on the approach or NN forces used, except for differences for
$N>28$ in the interactions among the $p$ orbitals, mostly
$p_{3/2}-p_{3/2}$ (due to more attractive second-order
core-polarization and third-order TDA/RPA contributions with $\vlowk$,
$\lm = 2.0 \fmi$, compared to the $G$-matrix). This uncertainty
remains when 3N forces are added [see Fig.~\ref{spe}~(c)]. Beyond
$^{48}$Ca, both NN-only theories predict a shell closure at $N=34$,
which is a major disagreement between the phenomenological models.

The dominant differences between phenomenological forces and NN-only
theory can be traced to their two-body monopole components, which
determine the average interaction between orbitals~\cite{PovesZuker,mono}.
In an operator expansion, the monopole interaction corresponds to the
term involving number operators, so that differences are enhanced with
$N$, 
and the SPE of orbital $l_j$ is effectively shifted by the monopole
interaction multiplied by the number of neutrons in orbital
$l'_{j'}$. The interplay of this two-body effect with the initial SPEs
largely determines the formation of shell structure.

Next, we include 3N forces among two valence neutrons and one nucleon
in the core. In Ref.~\cite{Oxygen}, we have shown that these
configurations give rise to repulsive monopole interactions among
excess neutrons. This corresponds to the normal-ordered two-body parts
of 3N forces, which was found to dominate in coupled-cluster
calculations~\cite{CC3N} over residual 3N forces (the latter should
be weaker because of phase space arguments~\cite{Fermi}). For
the $G$-matrix approach, we include FM 3N forces due to $\Delta$
excitations, 3N($\Delta$), where the parameters are fixed by standard
pion-N-$\Delta$ couplings~\cite{Green}. For chiral low-momentum
interactions, we take into account chiral 3N forces at
N$^2$LO~\cite{chiral3N}. These include long-range two-pion-exchange
parts $c_i$ (due to $\Delta$ and other excitations), plus
shorter-range one-pion exchange $c_D$ and 3N contact $c_E$
interactions that have been fit to the $^3$H binding energy and the
$^4$He matter radius~\cite{3Nfit}.
For $\vlowk$, we also consider the
one-$\Delta$ excitation 3N force that corresponds to particular values
for the two-pion-exchange part $c_i$ and $c_D=c_E=0$
[$\vlowk$+3N($\Delta$)]~\cite{chiral}.
In the $\vlowk$+3N calculations, we also scale all matrix
elements by $\hbar\omega \sim A^{-1/3}$.

For all results, full 3N multipole contributions are included to first
order~\cite{Kai}, although only the monopole part is responsible for
the SPE evolution in Fig.~\ref{spe}~(c). Here we see in both microscopic
approaches, 3N forces provide repulsive shifts of all single-particle
levels, changing the binding energies as shown in Fig.~\ref{gs}. In
addition, the repulsion between the $f_{7/2}$
and $p_{3/2}$ orbitals leads to an increased separation at $N=28$,
similarly to the phenomenological forces. Moreover, the gap at $N=32$ is
increased due to the repulsive $p_{3/2}-p_{1/2}$ interaction, while
$N=34$ remains approximately the same.

\begin{figure}
\begin{center}
\includegraphics[scale=0.48,clip=]{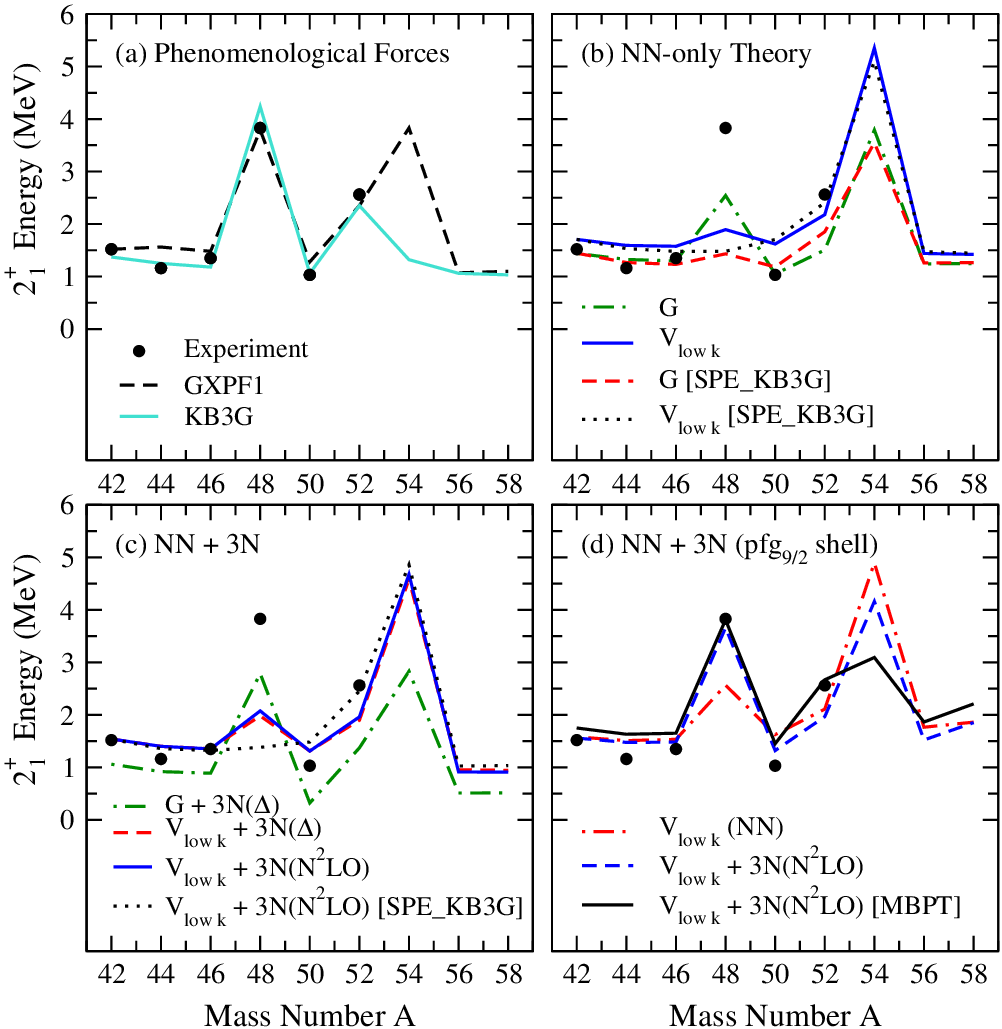}
\end{center}
\vspace*{-5mm}
\caption{First $2^+$ excitation energies in the even calcium isotopes
compared with experiment. (a)~Energies obtained from phenomenological 
models
KB3G~\cite{KB3G} and GXPF1~\cite{GXPF1}. 
(b)~NN-only theory:
energies based on a $G$-matrix and low-momentum interactions $\vlowk$
with empirical GXPF1 SPEs in $^{41}$Ca, as well as with KB3G values
[SPE\_KB3G]. (c)~Including contributions from 3N forces due to
$\Delta$ excitations, 3N($\Delta$), and chiral EFT 3N interactions at
N$^2$LO, 3N(N$^2$LO)~\cite{3Nfit}. (d)~Energies from 
$\vlowk$ and 3N(N$^2$LO) forces in the $pfg_{9/2}$ shell with
empirical GXPF1 SPEs and $g_{9/2}$ at $-1 \mev$ in $^{41}$Ca, as well
as with SPEs in $^{41}$Ca calculated consistently in MBPT.\label{2+}}
\vspace*{-4mm}
\end{figure}

We take into account many-body correlations by diagonalization in the
valence space and plot the first 2$^+$ energy of the even calcium
isotopes in Fig.~\ref{2+}. The excitation energies of the
phenomenological models in Fig.~\ref{2+}~(a) show the fit to the high
2$^+$ energy in $^{48}$Ca and hence the doubly-magic nature, and
highlight the difference in the prediction of $N=34$ as a shell
closure. In contrast, $^{48}$Ca is not reproduced in any calculation
based on NN forces in Fig.~\ref{2+}~(b), regardless of starting SPEs,
or whether we include the $g_{9/2}$ orbit in Fig.~\ref{2+}~(d).

\begin{figure}
\begin{center}
\includegraphics[scale=0.36,clip=]{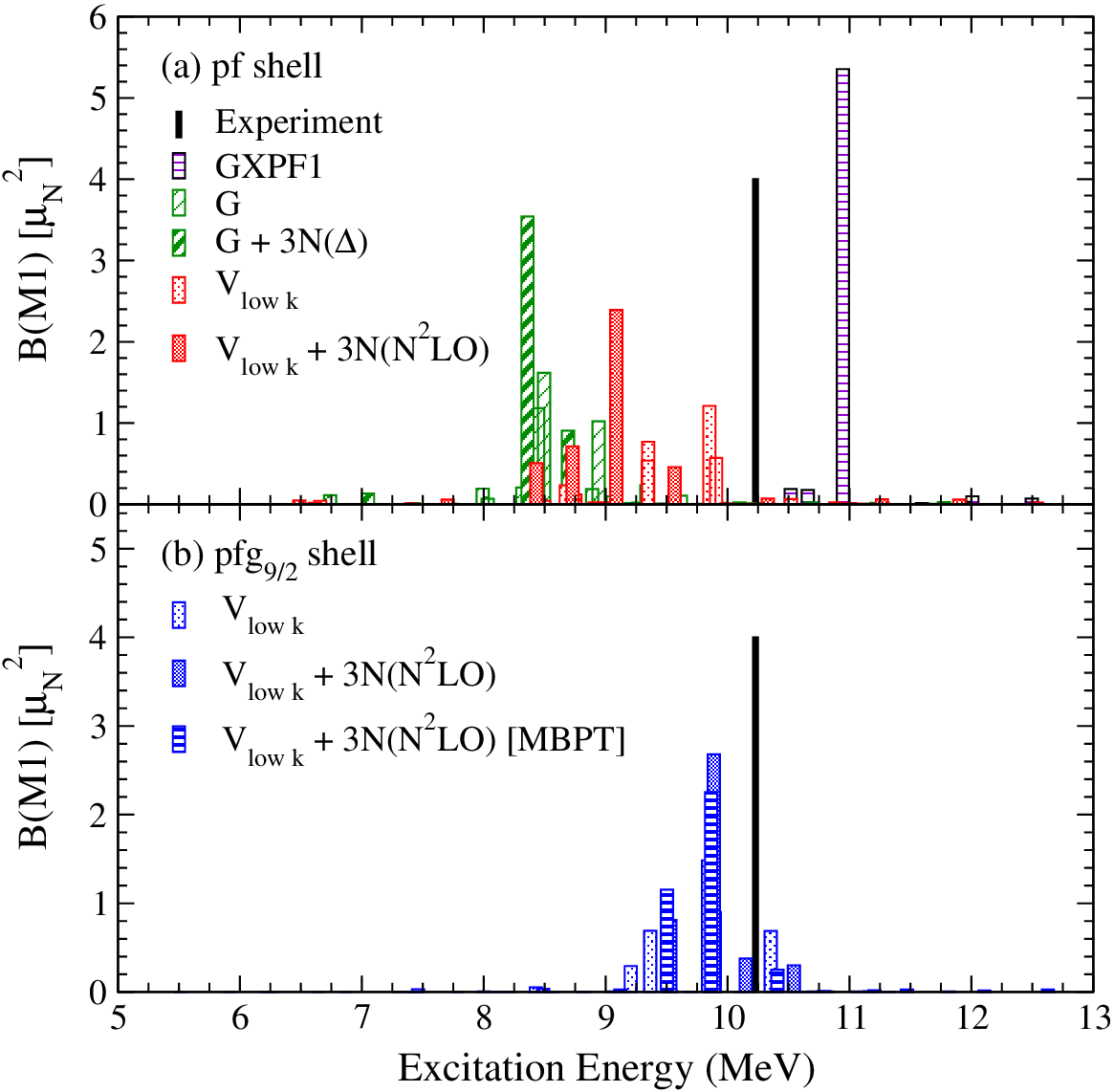}
\end{center}
\vspace{-5mm}
\caption{Magnetic dipole transition rates from the ground state to
1$^+$ excited states in $^{48}$Ca compared with
experiment~\cite{48CaM1}. The $B(M1)$ values are calculated in the
$pf$ and $pfg_{9/2}$ shell in panels~(a) and~(b), respectively, based
on NN-only interactions and including 3N forces (spin $g$ factors are
quenched by $0.75$). The results are labeled as in 
Fig.~\ref{2+}.\label{m1}}
\vspace*{-4mm}
\end{figure}

With 3N forces in Fig.~\ref{2+}~(c), the 2$^+$ energy in $^{48}$Ca is
uniformly improved. The $pf$ shell predictions with NN+3N forces are
similar with initial GXPF1 SPEs, but still below the experimental
value. With KB3G SPEs in $^{41}$Ca, the 2$^+$ energy is significantly
lower due to the smaller initial $f_{7/2}-p_{3/2}$ spacing. When the
$g_{9/2}$ orbit is included in Fig.~\ref{2+}~(d), the 2$^+$ energy is
obtained very close to experiment. In addition, we find that all
microscopic NN-only and NN+3N results at this level yield a high 2$^+$
energy in $^{54}$Ca, and hence a shell closure at N=34 (as suggested
in Ref.~\cite{magic}).
The similarities of
$\vlowk$+3N($\Delta$) and +3N(N$^2$LO) in Fig.~\ref{2+}~(c)
demonstrate that the configurations composed of valence neutrons probe
mainly the long-range parts of 3N forces.

To remove the uncertainty in the initial SPEs, we calculate the SPEs
in $^{41}$Ca by solving the Dyson equation, consistently including
one-body contributions to third order in MBPT in the same space as
the two-body interactions, and chiral 3N forces between one
valence neutron and two core nucleons to first order.
In contrast to the failure with NN-only forces, we find in
Fig.~\ref{spe}~(d) the $pf$ shell SPEs are generally similar to the
empirical ones, and we find the $g_{9/2}$ to initially lie between the
$p_{1/2}$ and $f_{5/2}$ orbitals. Our results based on MBPT SPEs and
consistent two-valence-neutron interactions are shown in
Fig.~\ref{2+}~(d). The agreement with experiment is very promising for
a parameter-free calculation based on NN and 3N forces. Furthermore,
the high 2$^+$ in $^{48}$Ca, despite a relatively small
$f_{7/2}-p_{3/2}$ gap, reflects the possible importance of correlations
beyond the $pf$-shell (in the context of SPEs, see also Ref.~\cite{CCSPE}).
Another challenge for microscopic
theories is the prediction of the first excited ($1/2^-$) state in
$^{49}$Ca, which indicates the size of the $p_{3/2}-p_{1/2}$ gap at
$N=28$. For $\vlowk$+3N(N$^2$LO) in both the $pf$ and $pfg_{9/2}$
shell, this energy is $\approx 1.0 \mev$ compared to the experimental
value $2.02 \mev$, while the MBPT results yield $1.8 \mev$.

We further examine the closed-shell nature of $^{48}$Ca in
Fig.~\ref{m1}, which shows the magnetic dipole transition rates
$B(M1)$ from the $0^+$ ground state to 1$^+$ excited states, where the
experimental concentration of strength indicates a single-particle
transition~\cite{48CaM1}. With NN-only forces, there is a significant
fragmentation of strength, and the energy of the dominant transition
is below the observed value. When 3N forces are included, the peak
energies are pushed up for all 3N(N$^2$LO) cases. Moreover, the MBPT
results predict a clear concentration, in very good agreement with
experiment.

\begin{figure}[t]
\begin{center}
\includegraphics[scale=0.48,clip=]{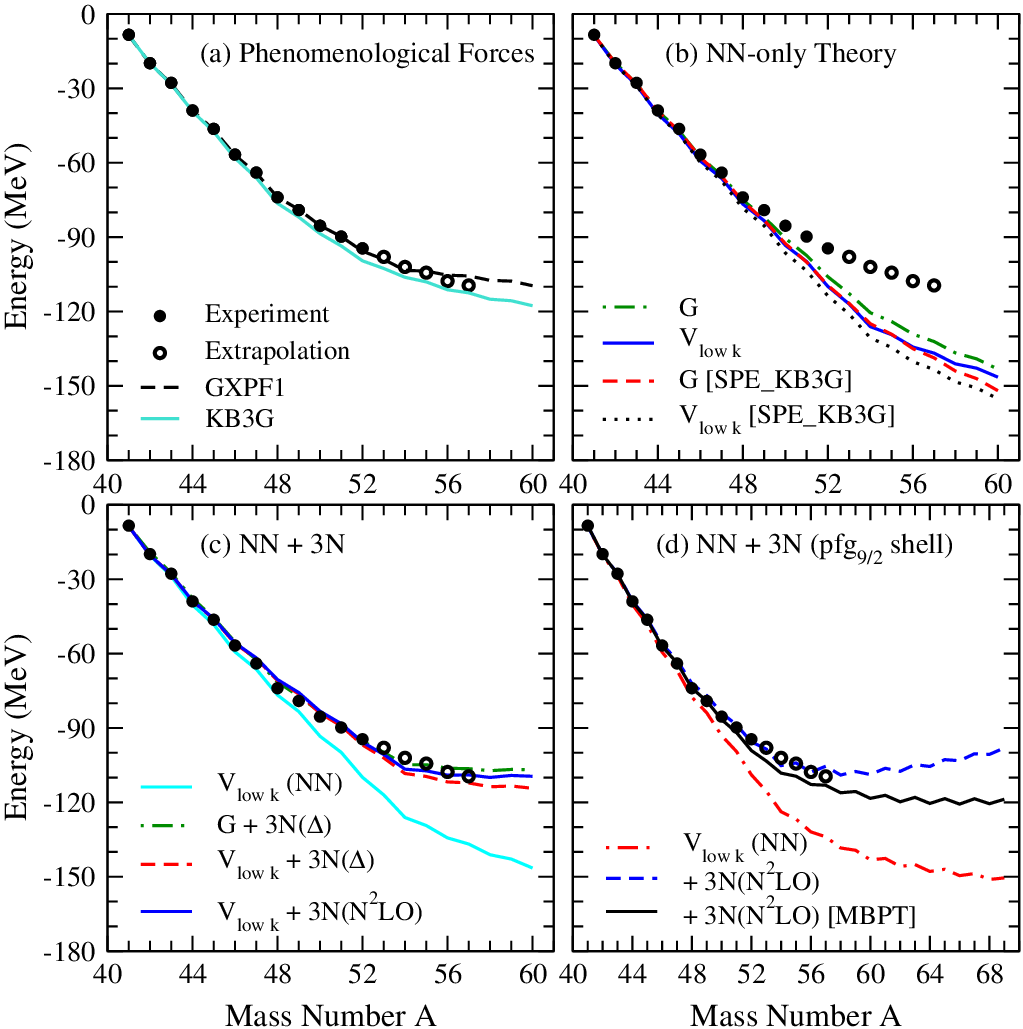}
\end{center}
\vspace{-5mm}
\caption{Ground-state energies of calcium isotopes relative
to $^{40}$Ca compared with experiment and extrapolated energies from
the AME2003 atomic mass evaluation~\cite{AME2003}. The panels and
results are labeled as in Fig.~\ref{2+}.\label{gs}}
\vspace{-4mm}
\end{figure}

Finally, we turn to the ground-state energies in Fig.~\ref{gs}, which
have been measured to $^{52}$Ca~\cite{AME2003} and are known to exist
to $^{58}$Ca~\cite{53-58Ca}. With phenomenological models, the
ground-state energies decrease to N=34, then the behavior flattens to
N=40 due to the weakly-bound $f_{5/2}$ orbital. With NN-only forces in
Fig.~\ref{gs}~(b) [as expected from Fig.~\ref{spe}~(b)], all
neutron-rich calcium isotopes are overbound. In Fig.~\ref{gs}~(c)
and~(d) the repulsion due to 3N forces leads to less bound
ground-state energies, and all calculations with 3N forces exhibit
good agreement with experiment [in Fig.~\ref{gs}~(c) the
$\vlowk$+3N(N$^2$LO) (KB3G SPE) results would lie on those of
$\vlowk$+3N($\Delta$) (GXPF1 SPE)].  The repulsive 3N mechanism,
discovered for the oxygen anomaly~\cite{Oxygen}, is therefore robust
and general for neutron-rich nuclei. In our best calculation with 
MBPT SPEs in the $pfg_{9/2}$
shell, the ground-state energies are modestly more bound.
Our results with
3N(N$^2$LO) suggest a drip line around $^{60}$Ca, which is close to
the experimental frontier~\cite{53-58Ca}. As the predicted energies
can significantly flatten from $N=34-40$, as is the case in our best MBPT 
calculation, the inclusion of continuum effects will be very important.

We have presented the first study of the role of 3N forces for binding
energies and evolution of shell structure in medium-mass nuclei, thus
linking the 3N forces frontier to the experimental frontier for
neutron-rich nuclei. Our results show that 3N forces and an extended
valence space are key to explain the $N=28$ magic number, leading to a
high $2^+$ excitation energy and a concentrated magnetic dipole
transition strength in $^{48}$Ca. It is intriguing and promising that
the parameter-free MBPT results in the extended valence space
reproduce experiment best. Future work will include a detailed
comparison to empirically adjusted interactions, where the $pfg_{9/2}$
interactions can also be transformed into $pf$-shell-only interactions
by an Okubo transformation.

\begin{acknowledgments}
This work was supported by the US DOE Grant DE-FC02-07ER41457 (UNEDF
SciDAC Collaboration) and DE-FG02-06ER41407 (JUSTIPEN), by 
grants-in-aid for Scientific Research~(A) 20244022 and~(C) 22540290,
the JSPS Core-to-Core program EFES, and the Alliance Program
of the Helmholtz Association (HA216/EMMI). Part of the numerical
calculations have been performed on Kraken at NICS, UT/ORNL, and at
the JSC, J\"ulich.
\end{acknowledgments}

\end{document}